\documentclass{sf2a-conf2022}
\usepackage{graphicx}
\usepackage{hyperref}
\usepackage[]{natbib}  
\usepackage{epstopdf}

\def\BibTeX{{\rm B\kern-.05em{\sc i\kern-.025em b}\kern-.08em
    T\kern-.1667em\lower.7ex\hbox{E}\kern-.125emX}}
\bibpunct{(}{)}{;}{a}{}{,}  

\begin{document}

\TitreGlobal{SF2A 2022}


\title{{\it Commission Femmes et Astronomie de la SF2A :}\\ Women participation in French Astronomy}

\runningtitle{Counting women}

\author{Rhita-Maria Ouazzani}\address{LESIA, Observatoire de Paris, Université PSL, Sorbonne Université, Université Paris Cité, CNRS, 5 place Jules Janssen, 92195 Meudon, France}
\author{Caroline Bot} \address{Université de Strasbourg, CNRS, Observatoire Astronomique de Strasbourg,
UMR7550, F--67000 Strasbourg, France}
\author{Sylvie Brau-Nogué} \address{IRAP, Universit\'e de Toulouse, CNRS, UPS, CNES, Toulouse, France}
\author{Danielle Briot}\address{Observatoire de Paris, 61 avenue de l’Observatoire, 75014 Paris, France}
\author{Patrick de Laverny}\address{Université Côte d'Azur, Observatoire de la Côte d'Azur, CNRS, Laboratoire Lagrange,F-06304 Nice, France}
\author{Nadège Lagarde} \address{Laboratoire d’Astrophysique de Bordeaux, Univ. Bordeaux, CNRS, B18N, allée Geoffroy Saint-Hilaire, 33615 Pessac, France}
\author{Nicole Nesvadba} \address{Université Côte d'Azur, Observatoire de la Côte d'Azur, CNRS, Laboratoire Lagrange,F-06304 Nice, France}
\author{Julien Malzac} \address{IRAP, Universit\'e de Toulouse, CNRS, UPS, CNES, Toulouse, France}
\author{Isabelle Vauglin}\address{Univ Lyon, Universit\'e Lyon1, ENS de Lyon, CNRS, CRAL UMR5574, F-69230 Saint-Genis-Laval, France}
\author{Olivia Venot}\address{Universit\'e de Paris Cit\'e and Univ Paris Est Creteil, CNRS, LISA, F-75013 Paris, France}




\setcounter{page}{237}


\maketitle


\begin{abstract}
The {\it Commission Femmes et Astronomie} conducted a statistical study that aims at mapping the presence of women in French professional Astronomy today, and set a starting point for studying its evolution with time. For the year 2021, we proceeded with a sub-set of 8 astronomy and astrophysics institutes, hosting a total of 1060 employees, among which PhD students, post-doctoral researchers, and academic, technical, and administrative staff, representing around 25\% of the community. We have investigated how the percentage of women vary with career stage, level of responsibility, job security, and level of income. The results of this preliminary study seem to illustrate the leaky pipeline, with one major bottleneck being the access to permanent positions. It appears that the proportion of women steadily decreases with the security of jobs, with the career stage, with the qualification level and with the income level. 

\end{abstract}



\section{Introduction}
The \textit{Commission Femmes et Astronomie} of the \textit{SF2A} (\textit{Société Française d'Astronomie et d'Astrophysique}, French astronomical Society) was created in 2020 to form an instance where questions related to gender equality can be addressed within the French astronomical community. The Commission has ten members, of which six members are currently --or were at some point-- also part of the SF2A Council: Caroline Bot, Sylvie Brau-Nogué, Danielle Briot*, Patrick de Laverny*, Nad\`{e}ge Lagarde*, Rhita-Maria Ouazzani*, Nicole Nesvadba, Julien Malzac*, Isabelle Vauglin, and Olivia Venot\footnote[1]{members of the SF2A Council}. The main goals of the Commission are to promote gender equality in Astronomy \& Astrophysics in France, fight against sexual and gender-based violence, support gender-focused outreach actions, \dots

Before the commission was created, different efforts to do a census of the status of women in astronomy in France were conducted. In particular, a survey was conducted through the SF2A to probe the future of doctors who obtained their PhD in Astronomy and Astrophysics between 2007 and 2017. The results presented in \citet{Berne2020} showed, among other points, that women were less likely to be offered permanent positions than men. That same year, \citet{Bot2020} did a census of the percentage of women on permanent positions in France, finding that 23\% percents of permanent positions at that time were held by women. Looking at two different age classes, they found that the number of women seemed to be decreasing for university positions while increasing for astronomer positions (CNAP) and that no evidence of a glass ceiling effect was observed. While both studies were important and necessary, they gave an instantaneous glimpse of the status of women in astronomy in 2019-2020, they were limited to the information requested or available and were biased by the surveyed population (young researchers for \citealt{Berne2020} or permanent positions for \citealt{Bot2020}). 

\begin{figure}[t!]
 \centering
\includegraphics[scale=0.5,clip]{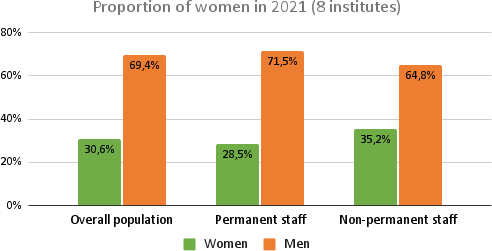}%
\includegraphics[scale=0.4,clip]{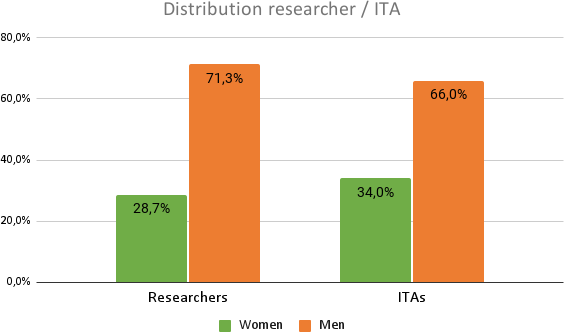}      
\caption{{\bf Left:} Proportion of women for the overall sample, among permanent staff, and non-permanent staff. {\bf Right:} Proportion of women among the researchers (from PhD to Emeritus, 687 individuals), and among administrative, technical and engineering staff (a.k.a ITA, 373 individuals). }
  \label{ouazzani_fig1}
\end{figure}

In this context, the {\it Commission Femmes et Astronomie} decided to conduct a statistical study that aims at mapping the presence of women in French professional Astronomy today, and set a starting point for studying its evolution with time.

\noindent As a first step, we would like to address general questions such as: 
\begin{itemize}
    \item What is the percentage of women in French Astronomical institutes?
    \item How does their number vary with their level of seniority --i.e. career stage--?
    \item What is the percentage of women at different levels of responsibility?
    \item How does the number of women depend on income level?
\end{itemize}

The perimeter of this study is restricted to the research units (institutes) depending of the Astronomy \& Astrophysics (AA) section of the National Institute for Universe Sciences (INSU), with the exception of the LISA institute, which was included in this study although it was not labeled AA, as part of this institute's activities are related to astronomy and astrophysics. The approach adopted consists in collecting data directly from the institutes heads, through their administrative services. The data is anonymised upfront, to comply with privacy policies. Once anonymised it is distributed to the members of the committee, who are the only persons authorized to manipulate them using secured tools. 

\noindent For the year 2021, taken as the starting point for the evolutionary sequence, we proceeded with a sub-set of institutes of the INSU-AA, as a proof of concept, with the aim of extending the study to all the INSU-AA institutes in the near future.

\section{Participation of women in Astronomy in 2021}

\subsection{The 2021 study set-up}

We were able to collect data from eight research institutes within France: GEPI, IRAP, Lagrange, LESIA, LISA, LUTh, ObAS and the SYRTE, which include 1060 individuals, representing around 25\% of the community. 

\noindent For all these institutes, the data contained the following entries: 
\begin{itemize}
    \item[-] Gender, 
    \item[-] Date of Birth, 
    \item[-] Employer (CNRS/CNAP/University/else), 
    \item[-] Status (students/post-doc/researcher/engineers, administrative or technical staff a.k.a. ITA),
    \item[-] and for the public servants: the category (known as {\it grade} in french: IR/IE/CR/DR/AA/A/MdC/PU/\dots).
\end{itemize}

\subsection{General results and job security}

The first number presented in Fig. \ref{ouazzani_fig1} (left, overall population), gives the proportion of women, regardless of their status, age or position. This number represents the likelihood of crossing paths with a woman in a corridor when walking through a French astronomy institute: one in three.

The workforce in French academia is composed of permanent staff, among which we count public servants --this includes persons in research, engineering, technical or administrative positions--,  as well as very few (but nevertheless growing number of) persons employed on corporate-like permanent positions. As for non-permanent positions, are counted PhD students, post-doctoral researchers, teaching assistants, apprenticeships, and holders of a short-term contract (on engineering, technical or administrative jobs).
Looking at the distribution of women among permanent and non-permanent positions (Fig.\ref{ouazzani_fig1}, left), we see that women (35.2\%) are most likely employed on temporary contracts than men (28.5\%). If we restrict this comparison to research positions, 25.0\% of researchers on permanent positions are women, whereas it is 34.9\% for non-permanent positions (28.7\% for the overall population).

One potential source of variability in these numbers is expected to come from the type of position (research or not). That is what is explored on the right panel of Fig.\ref{ouazzani_fig1}. We are aware that persons hired on ITA positions also contribute to the research that is produced in these institutes, but different social and economical values are attributed to research and ITA positions, and that is what, we believe, is determining here. Among women working in the 8 institutes included in the study, 39.2\% are hired as ITA, while for men the percentage is 33.4\%.

For the following discussions, we consider separately the population of researchers (in the broad sense: from PhD to Emeritus) on the one hand, and the population of Engineering, Technical and Administrative staff altogether (ITA) on the other hand.
\begin{figure}[!t]
 \centering
\includegraphics[scale=0.5,clip]{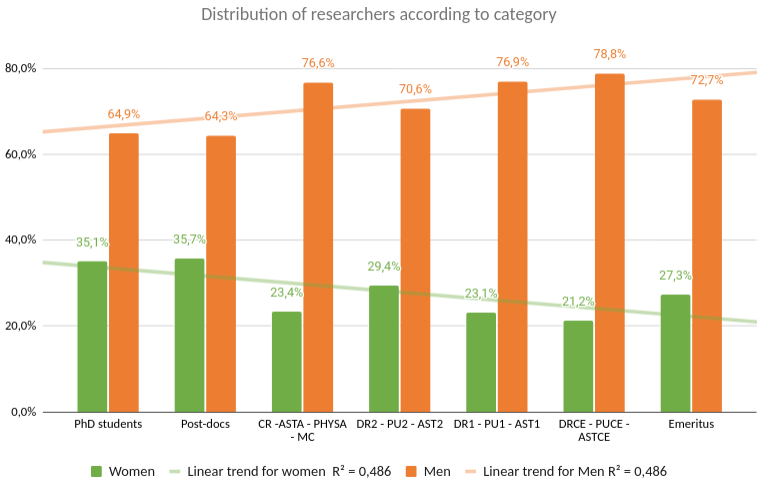}%
\caption{Proportion of women at each stage of the research career. In solid lines are superimposed linear fits of the proportion of women (light orange) and men (light green), the quality of the fit is indicated by a value of $R^2=0.486$.}
  \label{ouazzani_fig2}
\end{figure}
\subsection{Research career}
Concerning researchers, we have sorted the population (687 individuals) according to the category of their position. From the youngest to the most seniors, we have listed: PhD students, post-doctoral researchers, and positions equivalent to associate professors (CR, ASTA, PHYSA, MC), to {\it second-class} professors (DR2, PU2, AST2), to {\it first-class} professors (DR1, PU1, AST1), to professor of exceptional class (DRCE, PUCE, ASTCE), and Emeritus. 
The result is illustrated in Fig. \ref{ouazzani_fig2}. In order to emphasize the general trend, a linear fit has been performed ($R^2=0.486, p=0.08$), which shows that the proportion of women decreases with the career stage. This overall trend seems to be mostly caused by the first drop in the distribution: when women represent around 35\% of the population in non-permanent positions (PhD students and post-doctoral researchers), the proportion decreases to 23.4\% for the first level of permanent employment. After this first bottleneck, the number varies slightly between around 23\% and around 29\%, with two noticeable increases: one at the {\it second-class} professor level, and another one at the Emeritus level. 

\begin{figure}[!t]
 \centering
\includegraphics[scale=0.5,clip]{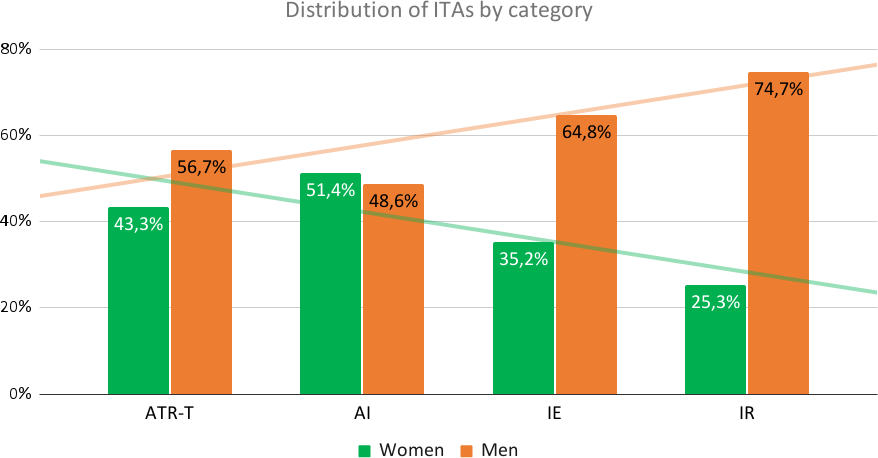}%
\caption{Proportion of women in each category of ITA jobs. From left to right they are ordered by qualification and income level. The solid lines give the optimal linear fits of these distributions, with a value of $R^2 =0.661$.}
  \label{ouazzani_fig3}
\end{figure}
It is important to bear in mind that this study gives a snapshot of the distribution for 2021. Temporal or causal relations between one stage to another are delicate to establish, and could properly be addressed only if this snapshot is renewed every year. However, concerning the clear increase of proportion of women between the associate professor level and the {\it second-class} professor level, one can wonder if it is not a stellar-main-sequence effect. Astronomers know that the large majority of observable stars are currently in their main sequence. That is because the main sequence, during which they burn hydrogen in their core, is the longest of all stellar evolution stages. Hence, we are led to wonder if this increase of women is not due to the fact that once they are promoted to {\it second-class} professorship, they spend a particularly long time in that stage before being promoted further up, if at all.

\subsection{ITA careers}
Concerning the population of ITA, composed here of 373 individuals, their distribution by career level is illustrated in Fig. \ref{ouazzani_fig3}. It is worth mentioning that the histogram presents broad categories of ITA (known as {\it corps} in French) ordered by levels of income and qualification, but contrary to Fig. \ref{ouazzani_fig2}, the progression from one category to the other is very little or none. Furthermore, the sources of variability are much more numerous than in the research career case. Firstly because of the variety of jobs it encompasses (administrative, technical, R\&D \dots). The jobs that fall into the ATR-T and AI categories have an dominant administrative component, whereas IE and IR are mostly scientific and technical jobs. Another source of variability are the qualifications needed to apply for these different kinds of positions: some require a PhD (IR), and others the High School Leaving Certificate ({\it Baccalauréat}).
In general, it is safe to assert that women are present in higher proportion in jobs that have a clear administrative component, and have lower levels of qualifications and lower income. As for men, they dominate in more technical jobs, where the level of qualification can be higher, and access higher income. In summary, we can conclude from Fig. \ref{ouazzani_fig3} that as the qualification and the level of income increases, the number of women decreases (illustrated by the linear fit, with a $p$ value of $p=0.1$). A finer analysis would require to blow up each of the histogram stick into finer category ({\it grade} in French), but with the current data at hand, we risk ending up with the small numbers statistics issue.

 \begin{figure}[t!]
 \centering
\includegraphics[scale=0.5,clip]{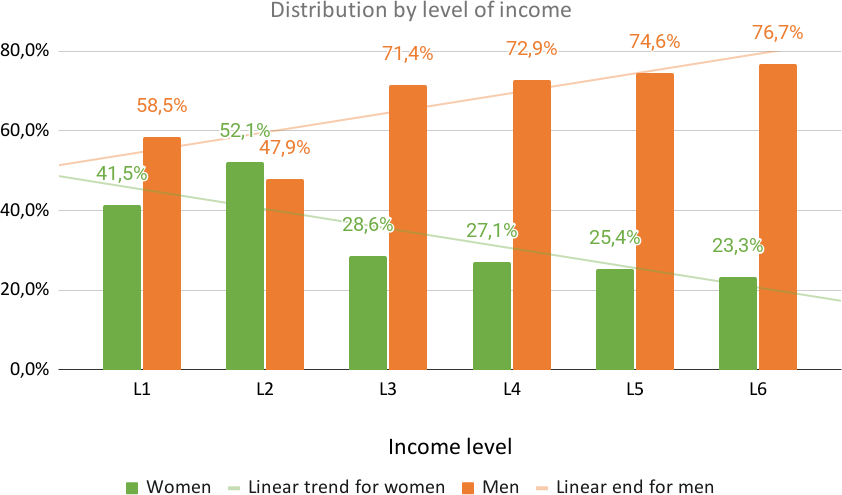}%
\caption{Proportion of women and men at each income level give in Table \ref{ouazzani_tab}, increasing from left to right. The quality of the fit is given by a value of $R^2=0.662$.}
  \label{ouazzani_fig4}
\end{figure}

\begin{table}[t!]
   \centering
    \begin{tabular}{|l|l|}
    \hline
        level & category \\
        \hline
        L1 & AJT, TCN, TCS\\
        L2 & TCE, AI\\
        L3 & IECN, IR2\\
        L4 & IEHC, IR1, CRCN, MC, PRAGCN, ASTA, PHYSA\\
        L5 & IRHC, CR1, CRHC, MCHC, DR2, PU2, AST2\\
        L6 & DR1, DRCE, PU1, PUCE, AST1, ASTCE, Emerites\\
        \hline
    \end{tabular}
    \caption{Scale of income for all the workers in the sample, researchers and ITA altogether. In terms of gross salary, it starts from about 1590 euros and reaches around 6200 euros.}
    \label{ouazzani_tab}
\end{table}

\subsection{The sinews of war}

We also addressed the crux of the matter: salary levels. For this purpose, we bring together again the whole sample (1060 individuals), and sort them out simply by level of income. To do so, we chose to use a scale of income set up by S. Brau-Nogué (see {\url{https://www.irap.omp.eu/egalite/bilan-social-et-parite-2022/}}) in her work to document gender inequalities in her institute (IRAP), work which has largely inspired this study. The scale is presented in Table \ref{ouazzani_tab}. Far from being a motivation for working in the academic sector, we believe that the level of income is a good indicator of the social value associated to a given position. Concerning the first two levels, they encompass mainly technical and administrative positions that are known to be positions which are very gender specific. But once we reach level L3, we observe that there is a clear and regular decrease of the proportion of women as the income increases. In solid lines are given the optimal linear fits of the distributions for men and women, with a $p$ value of $p=0.05$.

\section{Preliminary conclusions and perspectives}
We present the results of our first statistical study on the participation of women in French Astronomical Institutes. Our sample was composed of 1060 individuals, belonging to 8 institutes, making up for around 25\% of the targeted population. 
Although the results are still preliminary, it appears that the proportion of women steadily decreases with the security of jobs / the career stage / the qualification level / the income level. 
This seems to illustrate the well known leaky pipeline issue, but needs further confirmation.

In particular, the sample for this study suffered from a number of shortcomings. Some institutes included in the study cover topics which go beyond Astrophysics, such as the LISA and the SYRTE. Looking at the $p$ values associated to the trends determined, we wish to improve the statistical robustness of the inferences. We aim at solving this issue by extending the sample to all the French Astronomical Institutes in the coming years.
Moreover, the snap-shot nature of this study prevents from drawing strong conclusions about the evolutionary trends. Even if it can be very tempting to get a sense of evolution by looking at different generations of workers, one should keep in mind that state policies, or even the culture can change from one generation to the other.

\section*{Appendix: Index of jobs in A\&A}
Here we give some elements for understanding the zoo of jobs that one can find in the astronomical institutes in France. In general all the positions are divided into category (a.k.a. {\it corps} in french), and class (a.k.a. {\it grade} in french).\\

\noindent \underline{Academic careers:}
The three main job providers are the CNRS (French national institute for research), Universities, and the CNAP ({\it Conseil National des Astronomes et Physiciens}). 
According to the hiring institution, we can define different career paths, all of which contain 2 categories and each category can be split into 2 or 3 classes:
\begin{itemize}
\item CNRS: Chargés de Recherche (research associates; 2 classes: CRCN and CRHC) $\rightarrow$ Directeurs de Recherche (research directors; 3 classes: DR2, DR1, DRCE)
\item CNAP: Astronomes Associés (associate astronomers; 2 classes: ASTA, PHYSA) $\rightarrow$ Astronomes (astronomers; 3 classes: AST2, AST1, ASTCE)
\item University: Maitres de Conférence (associate professors, 2 classes: MC, MCHC) $\rightarrow$ Professeurs (professors; 3 classes: PU2, PU1, PUCE)
\end{itemize}

\noindent \underline{Engineering, technical, administrative careers (ITA):}
For ITA, the positions are divided into 5 categories, which are themselves divided into several classes (ordered by level of qualification): 
\begin{itemize}
    \item Adjoints techniques de la recherche (AJT)
    \item Techniciens de la recherche (TCN, TCS, TCE)
    \item Assistants ingénieurs (AI)
    \item Ingénieurs d'études (IECN, IEHC)
    \item Ingenieurs de recherche (IR2, IR1, IRHC)
\end{itemize}



\begin{acknowledgements}
The members of the \textit{Commission Femmes et Astronomie} would like to thank all institutes directors and colleagues who kindly agreed to providing the data that made this study possible, and the administrative staff who worked on their compilation and anonymisation.

\end{acknowledgements}

\bibliographystyle{aa}  


%
\end{document}